\documentstyle[preprint,aps,eqsecnum]{revtex}
\begin{document}
\def\bbuildrel#1_#2^#3{\mathrel{\mathop{\kern0pt #1}\limits_{#2}^{#3}}}
\draft
\preprint{LPTHE-Paris-95-06/DEMIRM-Obs. Paris-95007}

\title{\huge{\bf CLASSICAL SPLITTING OF FUNDAMENTAL STRINGS.}}

\vskip 2cm

\author{\Large{\bf  H.J. de Vega$^{(a)}$,  \\ J. Ram\'{\i}rez
Mittelbrunn$^{(b)}$,
M. Ram\'on Medrano$^{(b)}$,  \\ N. S\'anchez$^{(c)}$}}
\maketitle

\begin{center}

\bigskip

\bigskip

$^{(a)}$ Laboratoire de Physique Th\'eorique et Hautes Energies, Paris.\\
Laboratoire Associ\'e au CNRS UA 280,
Universit\'e Paris VI (P. et M. Curie)  et Universit\'e Paris VII
(D. Diderot).\footnote{Postal address: LPTHE, Tour 16, $1^{\rm er}$ \'etage,
        Universit\'e Paris VI,\\
        4, Place Jussieu, 75252, Paris cedex 05, FRANCE.}

\bigskip

\bigskip

 $^{(b)}$
Departamento de F\'{\i}sica Te\'orica, Madrid.\footnote{Postal
address: Facultad de Ciencias F\'{\i}sicas, Universidad
Complutense,  Ciudad Universtaria, \\ E-28040, Madrid, ESPA\~NA.}

\bigskip

\bigskip

$^{(c)}$ Observatoire de Paris, DEMIRM. \\
Laboratoire Associ\'e au CNRS UA 336, Observatoire de Paris et
\'Ecole Normale Sup\'erieure.\footnote{Postal address:
Observatoire de Paris, DEMIRM, 61, Avenue de l'Observatoire,
75014 Paris, FRANCE.}
\end{center}
\newpage
\begin{abstract}
We find exact solutions of the string equations of motion and constraints
describing the {\em classical}\  splitting of a string into two. We show that
for the same Cauchy data, the strings that split have {\bf smaller}
action than
the string without splitting. This phenomenon is already present in flat
space-time. The mass, energy and momentum carried out by the strings are
computed. We show that the splitting solution describes a natural
decay process of one string of mass $M$ into two strings with a
smaller total mass and some  kinetic energy. The standard
non-splitting solution is contained as a particular case. We also describe
the splitting of a closed string in the background of a singular
gravitational plane wave, and show how the presence of the strong
gravitational field increases (and amplifies by an overall factor) the
negative difference between the action of the splitting and non-splitting
solutions.

\end{abstract}
\newpage

\section{Introduction}

A great amount of work has been devoted to string theory in the lasts
years. However, very little attention has been paid to the theory of
fundamental strings as a classical theory. In particular, the
interaction of strings through joining and splitting has been only
considered at the {\em quantum } level. For example, the usual
procedure to compute the quantum string scattering amplitudes is based on
the evaluation of the correlation functions of vertex operators or
functionals, which are constructed out of a particular type of solution of
the classical string equations of motion and constraints: one in which the
string propagates without splitting and sweeping a world sheet that has the
topology of a cylinder or a strip \cite{Witten}.
In this paper, we will show that besides these classical solutions which
are used to build vertex operators, there also exist {\em classical}
solutions in which the string {\em splits}. More precisely, let us
consider as our classical action the Polyakov action \cite{Poly},
(we set here $2\alpha' = 1$ )
\begin{equation}
S  = \frac{1}{2\pi}\int_{W.S.} d\sigma d\tau \sqrt{-g} g^{ab} G_{AB}(X)
\partial_aX^A \partial_b X^B \label{1}
\end{equation}
\noindent
where $ g_{ab}$ is the world sheet metric and  $G_{AB}$ is the background
space-time metric. Since we are going to consider classical
solutions, both $ g_{ab}$ and $G_{AB}$  have Lorentzian signature. We
will show that there exists stationary points of $ S $ which
correspond to a classical splitting of one string into two. Moreover,
we will show that for some fixed Cauchy data $X^A(\sigma,\tau_0)$ and
${\dot{X}}^A(\sigma,\tau_0)$, the solutions corresponding to a string
that splits into two, have {\em smaller} action than the one in which the
string does not split, i.e. the sum of the areas swept by the two
pieces in which the string splits, is smaller than the area swept out by
the string that does not split. Of course,  the existence of
different kinds of solutions for the same Cauchy data is due to the
fact that the world sheet metric is not fixed by any dynamical
equation, and so we are free to choose it at will. Indeed, if we
enforce the conformal gauge {\em  globally} on our world sheet, we
are left with a world sheet that has the topology of a cylinder (for
closed strings), and therefore with a string that does not split.
However, nothing prevents us to consider a different world sheet
{\em topology}, and the interesting point is  that the solutions so
 obtained describe strings that split and have a {\em smaller} action
than the string that does not split. In order to show explicit
solutions of this type, we give here some interesting examples, saving
a more general discussion for future work. We will consider first an
example in flat space-time, to show how the phenomenon of smaller
action for the string that splits is already present in this
case. Secondly, we will consider a closed string moving in the
background given by a singular plane wave, and show how the  presence
of a strong gravitational field {\em increases} and {\em amplifies}
the negative difference between the action for the string that splits
and the action for the string that does not split. In fact, such
difference is amplified by an overall factor and  becomes infinitely
negative at the space-time singularity.
The classical splitting string solution contains the standard
(non-splitting) solution as a particular case. We also show that the
splitting solution describes a natural desintegration process of one string
of mass $M$ decaying into two pieces of a smaller total mass with some
kinetic energy.

It is  also possible to construct  string solutions where one string
splits into more than two pieces.

\section{String splitting in flat space-time}
\addtocounter{equation}{1}

Let us consider a closed string $X(\sigma,\tau)$ moving in a $ D$
dimensional flat Minkowski space-time. In order to describe a string
that splits, we choose a world sheet {\cal M} with the topology of a pant, and
call $(\sigma_0,\tau_0)$  the point at which the string breaks into two
pieces. To construct a solution for the string with this world sheet
topology, we consider in  {\cal M} the three regions $I$, $II$, and $III$
given by

\begin{equation}
\begin{array}{lcl}
I & \equiv & \{ (\sigma,\tau):~ 0\leq \sigma < 2\pi,~ \tau_i \leq \tau <
\tau_0 \}  \\
 & & \\
II & \equiv & \{ (\sigma,\tau):~ 0\leq \sigma < \sigma_0,~ \tau_0 < \tau
\leq \tau_f \} \\
 & & \\
III & \equiv & \{ (\sigma,\tau):~ \sigma_0 \leq \sigma < 2\pi,~ \tau_0 <
\tau \leq \tau_f \}
\end{array}
\label{2}
\end{equation}

\noindent
and impose the continuity of the dynamical variables $X(\sigma,\tau)$
and $\dot X (\sigma,\tau)$ at the splitting world sheet time $ \tau =
\tau_0 $

\begin{equation}
X^{(I)}(\sigma,\tau_0) = \left\{ \begin{array}{ll}
X^{(II)}(\sigma,\tau_0) & {\rm if}~ 0\leq \sigma <\sigma_0 \\
X^{(III)}(\sigma,\tau_0) & {\rm if}~ \sigma_0 \leq \sigma <2\pi
 \end{array} \right.
\label{3a}
\end{equation}
\begin{equation}
{\dot{X}}^{(I)}(\sigma,\tau_0) = \left\{ \begin{array}{ll}
{\dot{X}}^{(II)}(\sigma,\tau_0) & {\rm if}~ 0\leq \sigma <\sigma_0 \\
{\dot{X}}^{(III)}(\sigma,\tau_0) & {\rm if}~ \sigma_0 \leq \sigma <2\pi
\end{array} \right.
\label{3b}
\end{equation}
In each of the regions we are in the conformal
gauge. Thus, the equation of motion and the string constraints read the
same for the three functions $X^{(J)}(\sigma,\tau)$, with $J = I,II,III$
\begin{equation}
(\partial_{\tau}^2 - \partial_{\sigma}^2 ) X^{(J)} = 0 \label{4}
\end{equation}

\begin{equation}
({\dot{X}}^{(J)} \pm {X'}^{(J)})^2 = 0  \label{5}
\end{equation}

However, we must impose different boundary conditions for each of
the functions $X^{(J)}$. Since we want to describe a closed string
splitting into two closed strings, the appropriate boundary conditions
are the periodicity conditions

\begin{equation}
X^{(J)}(\sigma +\lambda_J, \tau) = X^{(J)}(\sigma , \tau )
\label{6}
\end{equation}

\noindent
where

\begin{equation}
\lambda_I = 2\pi \mbox{  }, \mbox{  } \lambda_{II}= \sigma_0 \mbox{
}, \mbox{  } \lambda_{III} = 2\pi - \sigma_0
\label{7}
\end{equation}

Of course, for the splitting to be possible, the string configuration
$X^{(I)}(\sigma,\tau)$ with which we start must satisfy the
consistency condition

\begin{equation}
X^{(I)}(0,\tau_0) = X^{(I)}(\sigma_0, \tau_0) = X^{(I)}(2\pi,\tau_0)
\label{8}
\end{equation}

The general solution of equations (\ref{4}) with the periodic
boundary conditions (\ref{6}) is

\begin{equation}
X^{(J)}(\sigma ,\tau ) = \sum_{n=-\infty}^{\infty} X_n^{(J)}(\tau)e^{i
\frac{2\pi}{\lambda_J} n\sigma} \label{9}
\end{equation}

\noindent
where

\begin{equation}
X_0^{(J)}= q^{(J)} + p^{(J)}\tau \label{10a}
\end{equation}

\noindent
and

\begin{equation}
X_n^{(J)}(\tau) = A_n^{(J)}\;e^{-i\frac{2\pi}{\lambda_J} n\tau} + B_n^{(J)} \;
e^{i\frac{2\pi}{\lambda_J} n\tau}~~{\rm for}~ n \neq 0
\label{10b}
\end{equation}

Now, in order to construct a solution of the equations of motion and
constraints corresponding to a string that splits, we begin with a
function $X^{(I)}(\sigma,\tau)$ that satisfies equations (\ref{4})
and the condition (\ref{8}) at $\tau = \tau_0 $, and then we
construct $X^{(II)}$ and $X^{(III)}$ by determining their Fourier
coefficients $X_n^{(II)}(\tau)$ and  $X_n^{(III)}(\tau)$ through the
matching conditions (\ref{3a}) and (\ref{3b}). Thus, from equations
(\ref{9}), (\ref{3a}) and (\ref{3b}) we obtain

\begin{equation}
\begin{array}{lcl}
X_n^{(II)}(\tau_0) &  = &   {\displaystyle i \sum_{m=-\infty}^{\infty}
\frac{1- e^{im\sigma_0}}{m\sigma_0 - 2\pi n} X_m^{(I)}(\tau_0) } \\
 & & \\
{\dot X}_n^{(II)}(\tau_0) & = & {\displaystyle i
\sum_{m=-\infty}^{\infty} \frac{1- e^{im\sigma_0}}{m\sigma_0 - 2\pi n}
{\dot X}_m^{(I)}(\tau_0) }
\end{array}
\label{11}
\end{equation}

\noindent
and

\begin{equation}
\begin{array}{lcl}
X_n^{(III)}(\tau_0) & = & {\displaystyle - ie^{- i\frac{2\pi}{2\pi-\sigma_0}
n \sigma_0} \sum_{m=-\infty}^{\infty} \frac{1- e^{im\sigma_0}}{m(2\pi-
\sigma_0 )  - 2\pi n} X_m^{(I)}(\tau_0) }  \\
 & & \\
{\dot X}_n^{(III)}(\tau_0) & = & {\displaystyle - ie^{-
i\frac{2\pi}{2\pi-\sigma_0} n \sigma_0} \sum_{m=-\infty}^{\infty} \frac{1-
e^{im\sigma_0}}{m(2\pi- \sigma_0 ) - 2\pi n} {\dot X}_m^{(I)}(\tau_0) }
\end{array}
\label{12}
\end{equation}

\noindent
so that equations (\ref{10a}), (\ref{10b}) and their
derivatives at $\tau = \tau_0 $ determine the constant coefficients
$(q,p,A_n,B_n)^{(II)}$ and $(q,p,A_n,B_n)^{(III)}$ in terms of the
initial data $(q,p,A_n,B_n)^{(I)}$. In this way we obtain the
solutions for the string pieces $II$ and $III$ from a general string
solution $I$. This last must satisfy the consistency condition
(\ref{8}).

It is important also to notice that the constraints (\ref{5}) are also
fulfilled by the strings $X^{(II)}$ and $X^{(III)}$. The matching conditions
(\ref{3a}) and (\ref{3b}) imply that both ${\dot X}(\sigma,Ê\tau )$
and $X'(\sigma,Ê\tau )$ are continuous at $\tau=\tau_0$. Therefore

\begin{equation}
\begin{array}{lclcl}
({\dot{X}}^{(II)} \pm {X'}^{(II)})^2(\sigma,Ê\tau_0 ) & = &
({\dot{X}}^{(I)} \pm {X'}^{(I)})^2(\sigma,Ê\tau_0 ) & = & 0 \\
 & & & & \\
({\dot{X}}^{(III)} \pm {X'}^{(III)})^2(\sigma,Ê\tau_0 ) & = &
({\dot{X}}^{(I)} \pm {X'}^{(I)})^2(\sigma,Ê\tau_0 ) & = & 0
\end{array}
\label{5p}
\end{equation}

\noindent
Then, taking into account the equation of motion (\ref{4}), it follows
that the constraints hold for all $\tau \geq \tau_0 $.

Let us now consider a particular example given by a circular
string that winds $r$ times upon itself:

\begin{equation}
\begin{array}{lcl}
T^{(I)} & = & M \tau  \\
& & \\
X^{(I)} & = & {\displaystyle \frac{M}{r}~ \sin r\tau~ \cos r\sigma } \\
& & \\
X^{(I)} & = & {\displaystyle \frac{M}{r}~ \sin r\tau~ \sin r\sigma }  \\
& & \\
X^{i(I)} & = & 0 ~~ {\rm for}~i= 3, \cdots, D-1
\end{array}
\label{13}
\end{equation}

\noindent
which satisfy

\begin{equation}
X^{(I)}(\sigma,0) = 0 = Y^{(I)}(\sigma,0) = T^{(I)}(0)
\label{14a}
\end{equation}

\noindent
and

\begin{equation}
{\dot{X}}^{(I)}(\sigma,0) = M\cos r\sigma~,~
{\dot{Y}}^{(I)}(\sigma,0) = M\sin r\sigma ~,~
{\dot{T}}^{(I)}(0) = M
\label{14b}
\end{equation}

It is inmediate to check that equations (\ref{13}) are a
solution of equation (\ref{4}) and satisfy the constraints (\ref{5}).

In order to obtain a {\em splitting} string solution, let us choose without
loss  of generality $\tau_0 = 0$. (One can always replace $\tau$ by
$\tau - \tau_0$ in equations (\ref{13})). Then the
splitting consistency condition (\ref{8}) is trivially fulfilled
because at $\tau = 0$ the string (\ref{13}) collapses
to a point.

For the string coordinate $T$, the matching conditions
(\ref{3a})-(\ref{3b}) yield

\begin{equation}
T^{(II)} = M\tau ~~, ~~T^{(III)} = M\tau   \label{15}
\end{equation}

Next, to obtain $X^{(II)}(\sigma,\tau)$, $Y^{(II)}(\sigma,\tau)$,
$X^{(III)}(\sigma,\tau)$ and $Y^{(III)}(\sigma,\tau)$, we observe that from
equations (\ref{13}) and (\ref{9})

\begin{equation}
\begin{array}{lclcl}
T^{(I)}(0) & = & T^{(I)}_0(0) & = & 0 \\
& & & & \\
{\dot{T}}^{(I)}(0) & = & {\dot{T}}^{(I)}_0(0) & = & M
\end{array}
\label{16a}
\end{equation}

\noindent
and

\begin{equation}
\begin{array}{lclcl}
X^{(I)}_n(0) & = & Y^{(I)}_n(0) & = & 0  \\
& & & & \\
{\dot{X}}^{(I)}_r(0) & = & {\dot{X}}^{(I)}_{-r}(0)
& = & {\displaystyle \frac{1}{2} M  }\\
& & & & \\
{\dot{Y}}^{(I)}_r(0) & = & -{\dot{Y}}^{(I)}_{-r}(0)
 & = & {\displaystyle  - \frac{i}{2} M } \\
& & & & \\
{\dot{X}}^{(I)}_n(0) &  = & {\dot{Y}}^{(I)}_n(0) & = & 0~~{\rm
for}~ n\neq \pm r
\label{16b}
\end{array}
\end{equation}

\noindent
Therefore, the matching equations (\ref{11}) and (\ref{12}) read
in this case

\begin{equation}
\begin{array}{lclcl}
X^{(II)}_n(0) & = & 0 & = & Y^{(II)}_n(0)   \\
& & & & \\
X^{(III)}_n(0) & = & 0 & = & Y^{(III)}_n(0)
\end{array}
\label{17a}
\end{equation}

\noindent
and

\begin{equation}
\begin{array}{lcl}
{\dot{X}}^{(II)}_n(0) & = & M \phi_n(\sigma_0)  \\
& & \\
{\dot{Y}}^{(II)}_n(0) & = & M \psi_n(\sigma_0)  \\
& & \\
{\dot{X}}^{(III)}_n(0) & = &  {\displaystyle M
e^{-i\frac{2\pi}{2\pi-\sigma_0}n\sigma_0} {\phi}^{\ast}_n(2\pi - \sigma_0) }
\\ & & \\
{\dot{Y}}^{(III)}_n(0) & = & {\displaystyle - M
e^{-i\frac{2\pi}{2\pi-\sigma_0}n\sigma_0} {\psi}^{\ast}_n(2\pi - \sigma_0)
}
\end{array}
\label{17b}
\end{equation}

\noindent
where

\begin{equation}
\begin{array}{lcl}
\phi_n(\sigma_0) & = & {\displaystyle - \frac{r\sigma_0\sin r\sigma_0 +
i\, 2\pi n \, (1 - \cos r\sigma_0)}{4{\pi}^2n^2 - r^2{\sigma_0}^2} } \\
& & \\
\psi_n(\sigma_0) & = & {\displaystyle - \frac{r\sigma_0 (1 - \cos r\sigma_0)
- i\, 2\pi n\, \sin r\sigma_0}{4{\pi}^2n^2 - r^2{\sigma_0}^2}  }
\end{array}
\label{18}
\end{equation}

Computing now the two sets of constants $(q,p,A_n,B_n)^{(II)}$ and
$(q,p,A_n,B_n)^{(III)}$  from expressions (\ref{10a}) and
(\ref{10b}), we obtain the time dependent Fourier coefficients
$X^{(II)}_n(\tau)$, $Y^{(II)}_n(\tau)$, $X^{(III)}_n(\tau)$ and
$Y^{(III)}_n(\tau)$. Finally the two pieces in which the string splits
read

\begin{equation}
\begin{array}{lcl}
X^{(II)}(\sigma,\tau) & = & {\displaystyle M\tau~ \frac{\sin
r\sigma_0}{r\sigma_0} +  M ~Ê\frac{\sigma_0}{2\pi}~ \sum_{n\neq 0}^{}
\frac{ \phi_n(\sigma_0)}{n}  \sin \left[ \frac{2\pi}{\sigma_0}n\tau \right]~
e^{i\frac{2\pi}{\sigma_0} n\sigma} }  \\
& & \\
Y^{(II)}(\sigma,\tau) & = & {\displaystyle M\tau ~ \frac{1 - \cos
r\sigma_0}{r\sigma_0}  + M ~\frac{\sigma_0}{2\pi}~ \sum_{n\neq 0}^{}
\frac{ \psi_n(\sigma_0)}{n}  \sin \left[ \frac{2\pi}{\sigma_0}n\tau  \right]
{}~e^{i\frac{2\pi}{\sigma_0} n\sigma} }
\end{array}
\label{19}
\end{equation}

\begin{equation}
\begin{array}{lcl}
X^{(III)}(\sigma,\tau) & = & {\displaystyle - M\tau ~
\frac{\sin r\sigma_0}{r(2\pi - \sigma_0)}
   +  M~ \frac{2\pi - \sigma_0}{2\pi}~ \sum_{n\neq 0}^{}
\left\{
\frac{1}{n}  \sin  \left[\frac{2\pi}{2\pi - \sigma_0}n\tau \right] \right. }
 \\ & & \\ & & \hspace{7cm}  {\displaystyle \left.
{\phi}^{\ast}_n(2\pi - \sigma_0)~ e^{i\frac{2\pi}{2\pi - \sigma_0}
n(\sigma - \sigma_0)} \right\} }   \\
& & \\
Y^{(III)}(\sigma,\tau) & = & {\displaystyle  - M\tau~ \frac{1 - \cos
r\sigma_0}{r (2\pi - \sigma_0)}
  - M~ \frac{2\pi - \sigma_0}{2\pi}
{}~\sum_{n\neq 0}^{} \left\{ \frac{1}{n}  \sin  \left[\frac{2\pi}{2\pi -
\sigma_0}n\tau \right] \right. }   \\
& & \\ & & \hspace{7cm}  {\displaystyle \left.Ê {\psi}^{\ast}_n(2\pi -
\sigma_0)~ e^{i\frac{2\pi}{2\pi - \sigma_0} n(\sigma- \sigma_0)} \right\} }
\end{array}
\label{20}
\end{equation}

Let us now discuss  the properties of the splitting string solution given by
equations (\ref{13}), (\ref{19}) and (\ref{20}). First, we notice that for
$n\neq 0$ and $\sigma_0 \rightarrow 2\pi $,

\begin{equation}
\begin{array}{lcl}
\phi_n(\sigma_0) & {\displaystyle \bbuildrel\longrightarrow_{ \sigma_0
\rightarrow 2\pi}^{} } & {\displaystyle \frac{1}{2}(\delta_{nr} +
\delta_{n(-r)}) }  \\    & & \\ & & \\
\psi_n(\sigma_0) & {\displaystyle \bbuildrel\longrightarrow_{ \sigma_0
\rightarrow 2\pi}^{} } &
- {\displaystyle \frac{i}{2}(\delta_{nr} - \delta_{n(-r)}) }
\end{array}
\label{21}
\end{equation}

\noindent
Therefore,

\begin{equation}
\begin{array}{lclcl}
X^{(II)}(\sigma,\tau)  & \bbuildrel\longrightarrow_{ \sigma_0
\rightarrow 2\pi}^{} & {\displaystyle \frac{M}{r}  \sin r \tau~
\cos r\sigma } & = & X^{(I)}(\sigma,\tau) \\
& & & & \\
Y^{(II)}(\sigma,\tau) & \bbuildrel\longrightarrow_{ \sigma_0 \rightarrow
2\pi}^{} &  {\displaystyle \frac{M}{r}  \sin r \tau ~Ê \sin r\sigma }
 &  =  & Y^{(I)}(\sigma,\tau)
\end{array}
\label{22}
\end{equation}

\noindent
Similarly for $\sigma_0 \rightarrow 0$,

\begin{equation}
\begin{array}{lcl}
X^{(III)}(\sigma,\tau) & \bbuildrel\longrightarrow_{ \sigma_0
\rightarrow 0}^{} &
X^{(I)}(\sigma,\tau) \\
& & \\
Y^{(III)}(\sigma,\tau) &
\bbuildrel\longrightarrow_{ \sigma_0Ê\rightarrow 0}^{}
 & Y^{(I)}(\sigma,\tau)
\end{array}
\label{23}
\end{equation}

Thus, the splitting string solution given by equations
(\ref{19}) and (\ref{20}) gives the solution
$X^{(I)}(\sigma,\tau)$ in the limits  $\sigma_0 \rightarrow 2\pi $ and
$\sigma_0 \rightarrow 0$  as it should be. In this sense, the  splitting
solution generalizes the standard string solution without splitting.

The energy and momentum carried out by each of the strings
$I,II$ and $III$ is given by

\begin{equation}
\begin{array}{lcl}
E^{(J)} & = & {\displaystyle \frac{1}{2\pi}
\int_0^{\lambda_J} \! d\sigma ~  {\dot{T}}^{(J)} } \\
& & \\
P_X^{(J)} & = & {\displaystyle \frac{1}{2\pi} \int_0^{\lambda_J} \! d\sigma
{}~{\dot{X}}^{(J)} }  \\
& & \\
P_Y^{(J)} & = & {\displaystyle \frac{1}{2\pi} \int_0^{\lambda_J} \! d\sigma
{}~{\dot{Y}}^{(J)} }
\end{array}
\label{24}
\end{equation}

\noindent
Then, using the Fourier series expansions (\ref{13}), (\ref{19}) and
(\ref{20}) we obtain

\begin{equation}
\begin{array}{lcl}
(E, P_X, P_Y)^{(I)} & = & M \,(1,0,0)  \\
  &  &  \\
(E, P_X, P_Y)^{(II)} & = &  {\displaystyle M \,(\frac{\sigma_0}{2\pi},
\frac{\sin r\sigma_0}{2\pi r}, \frac{1 - \cos r\sigma_0}{2\pi r}) } \\
  &  &  \\
(E, P_X, P_Y)^{(III)} & = &  {\displaystyle M\, (\frac{2\pi -
\sigma_0}{2\pi}, - \frac{\sin r\sigma_0}{2\pi r}, - \frac{1 - \cos
r\sigma_0}{2\pi r}) }
\end{array}
\label{25}
\end{equation}

{F}rom equations (\ref{25}) we see that the energy-momentum of the
string  before and after the splitting is conserved, as it should be. The
masses of the three strings are given by

\begin{equation}
\begin{array}{lcl}
M_I & = & M \\
 & & \\
M_{II} & = & {\displaystyle \frac{M}{2\pi} \sqrt{\sigma_0^2 -
\frac{4}{r^2} ~ {\sin }^2 \frac{r\sigma_0}{2} } }   \\
 & & \\
M_{III} & = & {\displaystyle \frac{M}{2\pi} \sqrt{ (2 \pi - \sigma_0)^2 -
\frac{4}{r^2} ~ {\sin }^2 \frac{r\sigma_0}{2} } }
\end{array}
\label{26}
\end{equation}

\noindent
Again, we see that

\begin{equation}
\begin{array}{lcl}
M_{II} & \bbuildrel\longrightarrow_{ \sigma_0
\rightarrow 2\pi}^{} & M_I  \\
 & & \\
M_{III} & \bbuildrel\longrightarrow_{ \sigma_0
\rightarrow 0}^{} & M_I
\end{array}
\label{27}
\end{equation}

\noindent
{}From equations (\ref{26}) it also follows that

\begin{equation}
 M_I \geq M_{II} + M_{III}
\label{28}
\end{equation}

\noindent
This tell us that the classical splitting string solution describes a
natural desintegration process, in which a string of mass $M$ decays into
two pieces with a smaller total mass and some kinetic energy, which depends
on  the point where the splitting takes place. Furthermore, one can see in
equations (\ref{25}), (\ref{19}) and (\ref{20}) that the two outgoing pieces
$II$ and  $III$ go away one from each other with opposite momenta.

	From equations (\ref{27}) it follows that the kinetic energy

\begin{equation}
K(\sigma_0) = M_I - (M_{II} + M_{III})
\label{29}
\end{equation}

\noindent
vanishes when

\begin{equation}
r \sigma_0 = 2l\pi~; ~l=0,1,\cdots, r
\label{30}
\end{equation}

This corresponds to cutting the initial string $X^{(I)}$ into two pieces,
$X^{(II)}$ and $X^{(III)}$, which contain an integer number of turns:
 $l$ and $r-l$ respectively. In this case, the equations (\ref{25}) tell us
that the momentum of the two pieces vanishes, and moreover the series
(\ref{19}) and (\ref{20}) sum up to $X^{(I)}(\sigma,\tau)$ and
$Y^{(I)}(\sigma,\tau)$ respectively. That is, cutting the string (\ref{13})
into two pieces which contain an integer number of circumferences, is
equivalent to not cutting it at all.( In other words,  a circular string
$X^{(I)}$ wound $r$ times is equivalent to two concentric strings
$X^{(II)}$ and  $X^{(III)}$ wound $l$ and $r-l$ times respectively). And
more generally, a circular string with $r=n$ turns is equivalent to r
strings with $n=1$ turn each. Of course, this is due to the fact that, in
this case, the periodicity conditions (\ref{6}) that we are enforcing for
$\tau > 0$ are already present in $X^{(I)}(\sigma, \tau)$  for $\tau<0$.

{}From equations (\ref{28}), (\ref{29}) and (\ref{30}) it follows that in
each of the $r$ equal windings in $\sigma$ described by the string $I$,
there is a splitting point $\sigma_0^i$, $i=1,2,\cdots r$ that maximizes
the kinetic energy of the pieces $II$ and $III$. That is, for each of the
turns of the string upon itself, there is an intermediate point where the
splitting of the string is most energetically favorable.

Also from eq. (\ref{26}) we see that the kinetic energy $K(\sigma_0)$ of
the pieces $II$ and $III$ decreases with the growing of $r$. In fact,
$K(\sigma_0) \rightarrow 0$ for $ r \rightarrow \infty$ and reaches its
maximum value for $r=1$ and $ \sigma_0 = \pi $. The value of this maximum
kinetic energy is

\begin{equation}
K_{{\rm max}} ={\displaystyle M \left( 1 -\sqrt{1-\frac{4}{\pi^2}} \right)}
\simeq 0.229 M
\label{kmax}
\end{equation}

\noindent
So, the most energetically favorable case correspond to a string with
$r=1$, which cuts into to equally long pieces. This also indicates that for
a string with $r>1$ the most energetically favorable process is not the
breaking of the string into two pieces, but the breaking into $r$ pieces at
the midpoints of each winding. Thus, the fundamental case to be considered
is that of a string with $r=1$ that cuts into two pieces.

Let us now discuss the string action. Let $S_I$ be the area swept by the
classical solution  $X^{(I)}$ that does not split, when it evolves in $\tau$
from $0$ to $\tau_f$; and $S_{II}$, $S_{III}$ the areas swept by the two
pieces of the splitting string evolving for the same $\tau$ interval. We are
interested in comparing  $S_I$ with $S_{II} + S_{III}$, for long enough
evolution time, i.e. $\tau_f \gg 2\pi$.

In flat space-time the string action (\ref{1}) takes the form

\begin{equation}
S = \frac{1}{2\pi} \int_0^{\tau_f} d\tau \int_0^{2\pi} d\sigma~ (-
{\dot{X}}^A {\dot{X}}_A + {X'}^A {X'}_A )
\label{31}
\end{equation}

\noindent
that using the constraints (\ref{5}), can be rewritten as

\begin{equation}
S = \frac{1}{\pi} \int_0^{\tau_f} d\tau \int_0^{2\pi} d\sigma
{}~Ê(\partial_{\sigma} X)^2
 \label{32}
\end{equation}

\noindent
Now for the string $I$, using the equation (\ref{13}) we have

\begin{equation}
S_{I} = \frac{1}{\pi} \int_0^{\tau_f} d\tau \int_0^{2\pi} d\sigma
{}~M^2 ~{\sin}^2 r\tau ~\sim ~ M^2 \tau_f ~~,~~(\tau_f \rightarrow \infty )
\label{33}
\end{equation}

\noindent
For the string $II$, using equations (\ref{15}) and (\ref{19}) we obtain

\begin{equation}
\begin{array}{lcl}
S_{II} & = & {\displaystyle \frac{1}{\pi} \int_0^{\tau_f} d\tau
\int_0^{\sigma_0}  d\sigma~
 \{ (\partial_{\sigma} X^{(II)})^2 + (\partial_{\sigma} Y^{(II)})^2 \} }\\
& = & {\displaystyle M^2~ \frac{\sigma_0}{\pi} \, \sum_{n \neq 0}^{}
\left( |\phi_n (\sigma_0) |^2 + |\psi_n (\sigma_0) |^2 \right)
\int_0^{\tau_f} d\tau~ {\sin}^2\, \left( \frac{2\pi}{\sigma_0}n\tau  \right)}
\end{array}
\label{34}
\end{equation}

\noindent
For large $\tau_f$ and $ n \neq 0$

\begin{equation}
\int_0^{\tau_f} d\tau ~ {\sin}^2  \left(\frac{2\pi}{\sigma_0}n\tau
\right) ~~\sim  ~
\frac{1}{2}\, \tau_f
\label{35}
\end{equation}

\noindent
On the other hand

\begin{equation}
\sum_{n \neq 0}^{}
\left( |\phi_n (\sigma_0) |^2 + |\psi_n (\sigma_0) |^2 \right) =
1- \frac{4}{r^2\sigma_0^2}~ {\sin}^2\,  \left(\frac{r\sigma_0}{2} \right)
\label{36}
\end{equation}

\noindent
Thus, for large $\tau_f$

\begin{equation}
S_{II} = M^2\tau_f ~\frac{\sigma_0}{2\pi}
\left[ 1- \frac{4}{r^2\sigma_0^2}~ {\sin}^2\,  \left( \frac{r\sigma_0}{2}
\right)\right]
\label{37}
\end{equation}

Notice that when $\sigma_0 \rightarrow 2\pi$, $S_{II} \rightarrow S_I$.
In addition, the use of the large $\tau_f$ approximation deserves the
following comment: one has to wait long enough time to appreciate the
difference between  the areas swept by the string $I$, and the strings $II$
and $III$. In fact, as it can be easily seen from equations (\ref{19}) and
 (\ref{20}), at first order in the Taylor expansion in $\tau$ around
$\tau = 0$, $X^{(I)}$, $X^{(II)}$ and $X^{(III)}$ coincide, and therefore
when $\tau_f \rightarrow 0$:

\begin{equation}
\frac{1}{\tau_f} (S_I - S_{II} - S_{III})
\bbuildrel\longrightarrow_{\tau_f \rightarrow 0}^{} 0
\label{38}
\end{equation}

Let us come back to the comparison between $S_{I}$ and $S_{II} + S_{III} $,
for large $\tau_f$. First, the behaviour of $ S_{III} $ is obtained from
equation (\ref{37}) through the replacement $\sigma_0 \rightarrow 2\pi -
\sigma_0 $

\begin{equation}
S_{III}~ \sim ~ M^2\tau_f ~ \frac{2\pi - \sigma_0}{2\pi}
\left[ 1- \frac{4}{r^2 (2\pi - \sigma_0)^2} ~ {\sin}^2\, \left(
\frac{r\sigma_0}{2}\right)  \right]
\label{39}
\end{equation}
\noindent
Then

\begin{equation}
S_{II} + S_{III} ~ \sim ~  M^2\tau_f \left\{ 1 - \frac{2}{\pi r^2}
\left( \frac{1}{\sigma_0} + \frac{1}{2 \pi - \sigma_0} \right)  {\sin}^2
\, \left(\frac{r\sigma_0}{2} \right) \right\}
\label{40}
\end{equation}

\noindent
and from equations (\ref{33}) and (\ref{40}) we obtain

\begin{equation}
\Delta S = (S_{II} + S_{III}) - S_{I} = - \,   M^2\tau_f~ \frac{2}{\pi r^2}
\left( \frac{1}{\sigma_0} + \frac{1}{2 \pi - \sigma_0} \right)
{\sin}^2 \, \frac{r\sigma_0}{2}
\label{41}
\end{equation}

\noindent
which is a negative quantity. Therefore, the string that splits sweeps a
smaller area than the string that does not split. From eq. (\ref{41}) it
also follows that the decrease in area $|\Delta S |$ vanishes for
$r\sigma_0  = 2l\pi$. This corresponds to the splitting into two pieces
containing an integer number of circumferences, which is equivalent
to non-splitting.

For the fundamental case $r=1$, the relative decrease in area is

\begin{equation}
\eta_1 (\sigma_0) = \frac{ |\Delta S_1|}{S_{1,I}} = \frac{2}{\pi}
\left( \frac{1}{\sigma_0} + \frac{1}{2 \pi - \sigma_0} \right)
{\sin}^2 \, \left(\frac{\sigma_0}{2} \right)
\label{eta1}
\end{equation}

which reaches its maximum value for $\sigma_0 = \pi$

\begin{equation}
\eta_{1,{\rm max}} = \frac{4}{\pi^2} \simeq 0.405
\label{eta1max}
\end{equation}

Thus, for the string configuration that we have chosen (eq. (\ref{13}) the
area swept by the strings that splits into two equally long pieces reduce
to a 60\% of the area swept by the string that does not split.

\section{String splitting in a singular plane wave background}
\addtocounter{equation}{50}

We discuss now the splitting solution for a closed string in a strong
gravitational field. We consider a $D$ dimensional singular gravitational
plane wave described by the metric

\begin{equation}
ds^2 = \frac{\alpha}{U^2}~ (X^2 - Y^2)\, dU^2  - dUdV + \sum_{j=2}^{D-1}
(dX^j)^2 \label{43}
\end{equation}

\noindent
where $U= X^0-X^1$, and $V= X^0+X^1$ are light cone coordinates,
$X^2\equiv X$, $X^3\equiv Y$, and $\alpha$ is a constant. This
space-time is sourceless and has  curvature on the null plane $U=0$.
 In this space-time, the classical string equations of motion
and constraints have been solved for the ordinary non-splitting string in
\cite{onda}. In this metric, the equation for $U$ is simply ${\partial }^2 U
= 0$. This allows us to take the light cone gauge exactly for all $\tau$

\begin{equation}
U = p\tau
\label{44}
\end{equation}

In this gauge the string equations for $X$ and $Y$ reduce to the linear
equations

\begin{equation}
\begin{array}{lcl}
{\displaystyle \left( -\partial_{\tau}^2 + \partial_{\sigma}^2 +
\frac{\alpha}{\tau^2} \right) X } & = & 0 \\
 & & \\
{\displaystyle \left( -\partial_{\tau}^2 + \partial_{\sigma}^2 -
\frac{\alpha}{\tau^2} \right) Y } & = & 0
\end{array}
\label{45}
\end{equation}

\noindent
which can be solved by Fourier expanding $X(\sigma,\tau)$ and
$Y(\sigma,\tau)$ in $\sigma$. Then, the $\tau$ dependent Fourier
 coefficients $X_n(\tau)$ and $Y_n(\tau)$ express in terms of
Bessel functions.

The remaining transverse coordinates $j = 4, \cdots,
D-1$ satisfy the flat space-time equations

\begin{equation}
\left( -\partial_{\tau}^2 + \partial_{\sigma}^2 \right) X^j  =  0~~;~~j =
4,\cdots , D-1
\label{46}
\end{equation}

Finally, the longitudinal coordinate $V$ is determined through the
constraints

\begin{equation}
G_{AB}~ \partial_{\pm} X^A~ \partial_{\pm} X^B = 0
\label{47}
\end{equation}

\noindent
which yield

\begin{equation}
\begin{array}{lcl}
p~ \partial_{\sigma} V & = & {\displaystyle 2 \sum_{j=2}^{D-1}
\partial_{\tau} X^j \partial_{\sigma} X^j } \\
 & & \\
p~ \partial_{\tau} V & = & {\displaystyle
 \frac{\alpha}{\tau^2}~(X^2-Y^2) + \sum_{j=2}^{D-1} \left\{
 (\partial_{\tau} X^j)^2 + (\partial_{\sigma} X^j)^2  \right\} }
\end{array}
\label{48}
\end{equation}

Let us describe now the splitting solution. We consider a generic string
configuration evolving in the region of negative $\tau$; i.e. before the
string reaches the singularity at $U=0$, and splitting at a certain point
$(\sigma_0,\tau_0)$ with $\tau_0 < 0$. We choose for the three strings
$ J = I, II, III$, a solution of the form

\begin{equation}
\begin{array}{lcl}
U^{(J)} & = & p\tau  \\
 & & \\
X^{(J)} & = & {\displaystyle \sum_{n=-\infty}^{\infty} X_n^{(J)}(\tau)\;
e^{i\frac{2\pi}{\lambda_J}n\sigma } } \\
 & & \\
Y^{(J)} & = & 0 \\
 & & \\
X^{j(J)} & = & 0 ~~,~~j = 4,\cdots, D-1
\end{array}
\label{49}
\end{equation}

\noindent
with the string coordinate $V$ determined through the constraints

\begin{equation}
\begin{array}{lcl}
p~ \partial_{\sigma} V^{(J)} & = & {\displaystyle 2~
\partial_{\tau} X^{(J)}~ \partial_{\sigma} X^{(J)} } \\
 & & \\
p~ \partial_{\tau}  V^{(J)} & = & {\displaystyle
 \frac{\alpha}{\tau^2}~Ê(X^{(J)})^2 + (\partial_{\tau} X^{(J)})^2
+ (\partial_{\sigma}X^{(J)})^2 }
\end{array}
\label{50}
\end{equation}

\noindent
The Fourier coefficients $X_n^{(J)}(\tau)$ are solutions of the equations

\begin{equation}
{\ddot{X}}_n^{(J)}  + \left[ {\left(\frac{2\pi n}{\lambda_J}\right)}^2 -
\frac{\alpha}{\tau^2} \right] X_n^{(J)} = 0
\label{51}
\end{equation}

\noindent
which can be written in the form

\begin{equation}
\begin{array}{lcl}
X_n^{(J)}(\tau) & = & {\displaystyle C_n^{(J)} \sqrt{-\tau}~ J_{-\nu} \left(
-\frac{2\pi}{\lambda_J} |n| \tau \right) +
D_n^{(J)} \sqrt{-\tau}~ J_{\nu} \left(
-\frac{2\pi}{\lambda_J} |n| \tau \right) }~,~ (n\neq 0) \\
 & & \\
X_0^{(J)}(\tau) & = & {\displaystyle C_0^{(J)}(-\tau)^{\frac{1}{2} - \nu} +
D_0^{(J)}(-\tau)^{\frac{1}{2} + \nu} }
\end{array}
\label{52}
\end{equation}

\noindent
where $ J_{-\nu}$ and $J_{\nu}$ are Bessel functions with index

\begin{equation}
\nu = \sqrt{\frac{1}{4} + \alpha}
\label{53}
\end{equation}

As in the flat space-time case, the functions $(U,V,X)^{(II)}$ and
$(U,V,X)^{(III)}$ which describe the evolution of strings $II$ and $III$,
are fixed by the initial string $(U,V,X)^{(I)}$, and the matching conditions
(\ref{3a}) and (\ref{3b}). However, since we are working now in the light
cone gauge, the following two remarks are in order. First, the choice of
 the light cone gauge for string $I$, and the matching conditions (\ref{3a})
and (\ref{3b})  for $U$, imply that the light cone gauge holds for the
pieces $II$ and $III$ as well, as stated in  eqs. (\ref{49}). Second, the
string coordinates $V^{(II)}$ and $V^{(III)}$ are determined through the
constraints (\ref{50}), instead through the matching conditions
(\ref{3a})-(\ref{3b}). However, this is consistent because the matching
conditions for the string coordinate $X$, together with the constraints
(\ref{3a})-(\ref{3b}) imply that $V^{(J)}$ also satisfy the matching
conditions (\ref{3a})-(\ref{3b}).

Let us choose now a specific initial configuration for the string coordinate
$X^{(I)}$ given by

\begin{equation}
X^{(I)} = {\displaystyle \frac{k}{r} f_{\nu} (\tau) \cos r\sigma  }
\label{54}
\end{equation}

\noindent
where

\begin{equation}
f_{\nu} (\tau) = r\sqrt{\tau \tau_0 } \left[ J_{\nu}(-r\tau_0 )
J_{-\nu}(-r\tau ) - J_{-\nu}(-r\tau_0 ) J_{\nu}(r\tau ) \right]
\label{55}
\end{equation}

This describes a straight string along the $X$-axis. According to eq.
(\ref{52}) we have chosen the constants $C_n^{(I)}$ and $D_n^{(I)}$ in the
form

\begin{equation}
\begin{array}{lclcl}
C_r^{(I)} & = & C_{- r}^{(I)} & = &{\displaystyle \frac{k}{2}~\sqrt{-\tau_0}
{}~J_{\nu}(-r\tau_0 ) }  \\
& & & & \\
D_r^{(I)} & = & D_{- r}^{(I)} & = &{\displaystyle \frac{k}{2}~\sqrt{-\tau_0}
{}~J_{-\nu}(-r\tau_0 ) } \\
& & & & \\
C_n^{(I)} & = & D_n^{(I)} & = & 0~~ {\rm for} ~ n\neq \pm r
\label{56}
\end{array}
\end{equation}

\noindent
This initial string configuration $(U,V,X)^{(I)}$ that we have chosen must
satisfy the splitting consistency condition (\ref{8}). For $U$
(eq.(\ref{44})) this condition is trivially satisfied. On the other hand,
eqs. (\ref{54})
 and (\ref{55}) yield

\begin{equation}
f_{\nu}(\tau_0 ) = 0
\label{57}
\end{equation}

\noindent
Thus

\begin{equation}
X^{(I)}(\sigma, \tau_0) = 0
\label{58}
\end{equation}

\noindent
and so the condition (\ref{8})  holds for the $X$ string coordinate.
Finally, using eqs. (\ref{54}) and (\ref{57}) in the constraints (\ref{50}),
we obtain

\begin{eqnarray}
p ~\partial_{\sigma} V^{(I)}(\sigma, \tau) &=& -k {\dot f}_{\nu}(\tau)
f_{\nu}(\tau) \; \sin(2 r \sigma) \cr
p ~\partial_{\sigma} V^{(I)}(\sigma, \tau_0) &=& 0
\label{60}
\end{eqnarray}

\noindent
i.e. $ V^{(I)}(\sigma, \tau_0)$ is independent of $\sigma$ and also
 satisfies the consistency condition (\ref{8}).

We can determine now the constants $(C_n^{(II)},D_n^{(II)})$ and
$(C_n^{(III)},D_n^{(III)})$ from the initial data $(C_n^{(I)},D_n^{(I)})$,
by using the matching conditions (\ref{3a}) and (\ref{3b}). These are

\begin{equation}
\begin{array}{lcl}
C_0^{(II)} & = & {\displaystyle K~\frac{\sin \nu \pi}{\nu \pi}
{}~(-\tau_0)^{\frac{1}{2} + \nu}~ \frac{\sin r \sigma_0}{r \sigma_0}}  \\
& & \\
C_n^{(II)} & = & {\displaystyle K~ \sqrt{-\tau_0}~ J_{\nu}\left(
-\frac{2\pi}{\sigma_0} |n| \tau_0 \right) \phi_n(\sigma_0)~,~ {\rm for}~
n\neq 0 } \\  & & \\
D_0^{(II)} & = &{\displaystyle - K~\frac{\sin \nu \pi}{\nu \pi}
{}~(-\tau_0)^{\frac{1}{2} - \nu} ~\frac{\sin r \sigma_0}{r \sigma_0}} \\
& & \\
D_n^{(II)} & = &{\displaystyle - K~ \sqrt{-\tau_0}~ J_{-\nu}\left(
-\frac{2\pi}{\sigma_0} |n| \tau_0 \right) \phi_n(\sigma_0) ~,~ {\rm for}~
n\neq 0 }
\end{array}
\label{61}
\end{equation}

\noindent
and

\begin{equation}
\begin{array}{lcl}
C_0^{(III)} & = &{\displaystyle - K~\frac{\sin \nu \pi}{\nu \pi}
{}~(-\tau_0)^{\frac{1}{2} + \nu}~ \frac{\sin r \sigma_0}{r \sigma_0} } \\
& & \\
C_n^{(III)} & = &{\displaystyle K ~e^{-i \frac{2\pi}{2\pi-\sigma_0}n\sigma_0}
{}~\sqrt{-\tau_0}~ J_{\nu}\left( -\frac{2\pi}{2\pi-\sigma_0} |n| \tau_0 \right)
{\phi_n}^{\ast}(2\pi -\sigma_0)~,~ {\rm for}~ n\neq 0 }\\
& & \\
D_0^{(III)} & = & {\displaystyle K~\frac{\sin \nu \pi}{\nu \pi}
{}~(-\tau_0)^{\frac{1}{2} - \nu} ~\frac{\sin r \sigma_0}{r \sigma_0} } \\
& & \\
D_n^{(III)} & = &{\displaystyle - K ~ e^{-i
\frac{2\pi}{2\pi-\sigma_0}n\sigma_0}
 ~\sqrt{-\tau_0}~ J_{-\nu}\left( -\frac{2\pi}{2\pi-\sigma_0}
|n| \tau_0 \right) {\phi_n}^{\ast}(2\pi -\sigma_0) ~,~ {\rm for}~ n\neq 0
}
\end{array}
\label{62}
\end{equation}

Hence, the Fourier expansions (\ref{49}) for the $X$ coordinates of strings
$II$ and $III$ read

\begin{equation}
\begin{array}{lcl}
X^{(II)}(\sigma, \tau) & = &{\displaystyle X_0^{(II)}( \tau) + }\\
& & {\displaystyle
\frac{\sigma_0}{2\pi}~ k~  \sum_{n \neq 0}^{} \left\{ \frac{1}{|n|}
 ~F_{\nu}\left( \frac{2\pi}{\sigma_0} |n| \tau ,\frac{2\pi}{\sigma_0} |n|
\tau_0 \right) \phi_n(\sigma_0)~
e^{i\frac{2\pi}{\sigma_0} n\sigma} \right\} }
\end{array}
\label{63}
\end{equation}
\noindent
and
\begin{equation}
\begin{array}{lcl}
X^{(III)}(\sigma, \tau)  =  {\displaystyle X_0^{(III)}( \tau)}  \\
+ {\displaystyle \frac{2\pi
-\sigma_0}{2\pi}~ k~  \sum_{n \neq 0}^{} \left\{ \frac{1}{|n|}~ F_{\nu}\left(
\frac{2\pi}{2\pi -\sigma_0} |n| \tau ,\frac{2\pi}{2\pi -\sigma_0} |n| \tau_0
\right)
{\phi_n}^{\ast}(2\pi -\sigma_0)
{}~e^{i\frac{2\pi}{2\pi -\sigma_0} n(\sigma -\sigma_0)} \right\} }
\end{array}
\label{64}
\end{equation}

\noindent
where

\begin{equation}
F_{\nu}(u,v) = \sqrt{uv} \left[ J_{\nu}(-v) J_{-\nu}(-u)
 - J_{-\nu}(-v) J_{\nu}(-u) \right]
\label{65}
\end{equation}

\noindent
and

\begin{equation}
X_0^{(II)}( \tau)  = - X_0^{(III)}( \tau) =  k~ \frac{\sin \nu \pi}{\nu \pi}
{}~\frac{\sin r \sigma_0}{r \sigma_0}~ \sqrt{\tau_0 \tau}
\left[ \left( \frac{\tau_0}{\tau} \right)^{\nu} - \left( \frac{\tau}{\tau_0}
\right)^{\nu} \right]
\label{66}
\end{equation}

Let us discuss now the properties of the splitting string solution given by
(\ref{63}) and (\ref{64}). First, we notice that from eq. (\ref{21})

\begin{equation}
\begin{array}{lcl}
X^{(II)}(\sigma, \tau) & \bbuildrel\longrightarrow_{\sigma_0 \rightarrow 2
\pi }^{} & X^{(I)}(\sigma, \tau) \\ & & \\
X^{(III)}(\sigma, \tau) & \bbuildrel\longrightarrow__{\sigma_0
\rightarrow 0 }^{} & X^{(I)}(\sigma, \tau)
\end{array}
\label{67}
\end{equation}

That is, the splitting solution  with strings $II$ and $III$ contains the
one string non-splitting solution as a particular case. In addition,  for

\begin{equation}
r\sigma_0 = 2l\pi ~~ ;~~ l = 1, \cdots, r-1
\label{68}
\end{equation}

\noindent
the Fourier series (\ref{63}) and (\ref{64}) sum up to
$X^{(I)}(\sigma, \tau)$. Thus again, cutting the string by an integer number
of windings is equivalent to not cutting it at all.

 Let us study now the string action. We want to compare the area $S_I$ swept
by the string without splitting, with the areas $S_{II}$ and $S_{III}$ swept
by the strings $II$ and $III$. We consider the evolution of the three
strings for the same  $\tau$ interval

\begin{equation}
\tau_0 \leq \tau \leq \tau_f < 0
\label{69}
\end{equation}

\noindent
in the ingoing region, where the string has not yet reached the
singularity at $\tau = 0$. We shall compute the action
for a long $\tau$ interval, i.e. $\tau_f - \tau_0 \gg 2\pi $, as we did in
flat space-time. However, in this case we shall implement this
approximation by letting $\tau_0 \rightarrow -\infty$, and allowing
$\tau_f \rightarrow 0^-$
in order
to incorporate the effect of the space-time singularity at $\tau = 0$.

In the space-time (\ref{43}) the string action is

\begin{equation}
S = \int \! \int_{W.S.} d\tau d\sigma \left\{ \frac{\alpha}{U^2} (X^2 - Y^2)
\, \partial_a U \partial^a U - \partial_a U \partial^a V +
\sum_{j=2}^{D-1} \partial_a X^j \partial^a X^j \right\}
\label{70}
\end{equation}

In the light cone gauge, using the constraints (\ref{50}) and for the
particular string configuration (\ref{49}), the action for the three
strings takes the form

\begin{equation}
S_J = \frac{1}{\pi} \int_{\tau_0}^{\tau_f} d\tau \int_{0}^{\lambda_J}
 d\sigma ~(\partial_{\sigma} X^{(J)})^2
\label{71}
\end{equation}

\noindent
Then, using eqs. (\ref{54}),(\ref{63}) and (\ref{65}), we have

\begin{equation}
S_I = k^2 \int_{\tau_0}^{\tau_f} d\tau ~F_{\nu}^2 (r\tau ,r\tau_0 )
\label{72}
\end{equation}

\noindent
and

\begin{equation}
S_{II} = k^2~ \frac{\sigma_0}{\pi} \int_{\tau_0}^{\tau_f} d\tau ~
\sum_{n \neq 0}^{} ~| F_{\nu} \left( \frac{2\pi}{\sigma_0} |n| \tau,
 \frac{2\pi}{\sigma_0} |n| \tau_0 \right) |^2 ~|\phi_n(\sigma_0)|^2
\label{73}
\end{equation}

\noindent
{}From eq. (\ref{22}) we see that $S_{II} \rightarrow S_I $ when $ \sigma_0
\rightarrow 2\pi $ as it should be.

We shall do the comparison of the two areas $S_I $ and $S_{II} + S_{III}$
in two regimes: first for $\alpha = 0$ which corresponds to flat space-time,
and then for $\alpha \geq 3/4$ which correspond to a strong gravitational
wave.

For $\alpha = 0 $, the index $\nu $ is $1/2$ and the Bessel functions
reduce to circular functions:

\begin{equation}
F_{1/2} \left( \frac{2\pi}{\sigma_0} |n| \tau, \frac{2\pi}{\sigma_0} |n|
\tau_0 \right) = \frac{2}{\pi} \sin \left( \frac{2\pi}{\sigma_0} |n|
(\tau - \tau_0) \right)
\label{74}
\end{equation}

\noindent
Then, for $\tau_0 \rightarrow -\infty $, eqs. (\ref{72}) and (\ref{73}) yield

\begin{equation}
 S_I^{(\alpha = 0)} = \frac{4k^2}{\pi^2} \int_{\tau_0}^{\tau_f} d\tau
{}~{\sin }^2\, r (\tau -\tau_0) \sim \frac{2k^2}{\pi^2}(\tau_f -\tau_0)
\label{75}
\end{equation}

\noindent
and

\begin{equation}
\begin{array}{lcl}
 S_{II}^{(\alpha = 0)} & = & {\displaystyle \frac{4k^2}{\pi^3} ~ \sigma_0
\int_{\tau_0}^{\tau_f} d\tau ~ \sum_{n \neq 0}^{} {\sin }^2 \left(
\frac{2\pi}{\sigma_0} |n|  (\tau - \tau_0) \right) } ~
|\phi_n(\sigma_0)|^2 \;
\sim \\
& &  \\
& \sim & {\displaystyle \frac{2k^2}{\pi^2} ~ (\tau_f - \tau_0) ~
\frac{\sigma_0}{\pi} ~ \sum_{n \neq 0}^{} |\phi_n(\sigma_0)|^2 } ~ = \\
& &  \\
& = & {\displaystyle \frac{2k^2}{\pi^2}~ (\tau_f - \tau_0) ~
\frac{\sigma_0}{2\pi}
 \left( 1 +  \frac{\sin 2r\sigma_0 }{2r\sigma_0} -
\frac{2\, {\sin}^2\, r\sigma_0 }{r^2\sigma_0^2} \right)  }
\end{array}
\label{76}
\end{equation}

\noindent
In addition, replacing $ \sigma_0 \rightarrow 2\pi - \sigma_0 $, we have
\begin{equation}
 S_{III}^{(\alpha = 0)} \sim \frac{2k^2}{\pi^2} ~(\tau_f - \tau_0)
{}~ \frac{2\pi - \sigma_0}{2\pi} \left[ 1 -
\frac{\sin 2r\sigma_0 }{2r(2\pi - \sigma_0)}
- \frac{2\, {\sin}^2\, r\sigma_0}{r^2(2\pi - \sigma_0)^2} \right]
\label{77}
\end{equation}

\noindent
Thus

\begin{equation}
{\Delta S}^{(\alpha = 0)} = (S_{II} + S_{III}) - S_{I} = -
 \frac{1}{\pi r^2}
\left( \frac{1}{\sigma_0} + \frac{1}{2\pi - \sigma_0} \right)
{\sin}^2 r\sigma_0 ~S_{I}^{(\alpha = 0)}
\label{78}
\end{equation}

\noindent
which is  a negative quantity. Again, the string that splits has smaller
action than the non-splitting one. However the result (\ref{78}) is
different from (\ref{41}), because we have here a different string
configuration. In particular, the action difference (\ref{78}) vanishes in
the present case for

\begin{equation}
r\sigma_0 = (2s + 1) \pi~~,~~ s = 0,1, \cdots, \left[ \frac{r-1}{2} \right]
\label{80}
\end{equation}

\noindent
i.e. it vanishes not only for an integer number of windings $~ r\sigma_0 =
2l\pi,~l=0,1,\cdots, r $, but also for a half integer number of windings.
This is so because the barycentric term (eq.(\ref{66})) vanish  for an
integer or half integer number of windings. However, the Fourier expansions
(\ref{63}) and (\ref{64}) sum up to $X^{(I)}(\sigma,\tau)$ for an integer
number of windings, but do not for a half integer number of windings. This
happens here because when $\sigma_0$ corresponds to a half integer number
of windings, we have a straight string configuration with $X'(0,\tau_0) =
X'(\sigma_0,\tau_0) = 0 $. Hence the initial closed string may split into
two open strings. Thus, in this case the strings $II$ and $III$ are open
strings that stay together and change their respective shapes compared with
string $I$.

For the fundamental case $r=1$, the relative decrease in area is

\begin{equation}
\eta_1^{(\alpha =0)}(\sigma_0) = \frac{ |\Delta S_1|}{S_{1,I}} =
\frac{1}{\pi} \left( \frac{1}{\sigma_0} + \frac{1}{2\pi - \sigma_0} \right)
{\sin}^2 \sigma_0
\label{eta2}
\end{equation}

which has two symmetric maxima at the points $\sigma_0 = 1.291 $ and
$\sigma_0 = 2\pi -1.291$ with heigth

\begin{equation}
\eta_{1,{\rm max}} = 0.287
\label{eta2max}
\end{equation}

Notice that in this case $\sigma_0 = \pi$ gives a minimum with
$\eta_1(\pi)$ = 0, corresponding to the splitting of the closed string into
two open strings.

Let us turn now to the discussion of the regime $\alpha \geq 3/4$, ($ \nu
\geq 1$). In this case, the integral (\ref{72}) diverges in the
limit $\tau_f \rightarrow 0^- $. Thus, for $\nu \geq 1$
the behaviour of
 $S_I$ for $\tau_f \rightarrow 0^-$ is dominated by the upper
limit in the integral (\ref{72}). That is, the most important contribution
to $S_I$
 comes from the region near the singularity. It is in this sense that we
talk of a strong enough gravitational wave for $\alpha \geq 3/4$. For
$\tau_f \rightarrow 0^- $ we have

\begin{equation}
S_I \sim k^2~Ê\frac{2^{2\nu}~Ê \tau_0}{\Gamma^2 (1-\nu)~(2-2\nu )~ r^{2\nu
-2}}~J_{\nu}^2 (-r \tau_0) \left( \frac{\tau_f}{\tau_1}Ê\right)^{2-2\nu}
\label{81}
\end{equation}

\noindent
where $\tau_1$ is an intermediate point in the interval $\tau_0 <
\tau_1 < \tau_f  < 0$, and we assume that $\tau_0$ is such that

\begin{equation}
J_{\nu}^2 (-r \tau_0) \neq 0
\label{82}
\end{equation}

Let us consider the series in the integrand of eq. (\ref{73}). The terms
with $|n\tau | \ll 1 $ behave as

\begin{equation}
\begin{array}{l}
{\displaystyle | F_{\nu} \left( \frac{2\pi}{\sigma_0} |n| \tau,
 \frac{2\pi}{\sigma_0} |n| \tau_0 \right) |^2 \sim } \\  \\
\sim
{\displaystyle \frac{2^{2\nu}}{\Gamma^2(1-\nu)} \left( \frac{2\pi
|n|}{\sigma_0} \right)^{2-2\nu}(-\tau_0)~ J_{\nu}^2 \left( -\frac{2\pi
}{\sigma_0} |n| \tau_0 \right) (-\tau)^{1-2\nu} }
\end{array}
\label{83}
\end{equation}

\noindent
For $|n \tau | \geq 1$, and small enough $|-\tau|$ (i.e. $ |n| \gg 1$), the
terms in the series of eq. (\ref{73}) are very much suppressed because
 for $n \rightarrow \infty$, $|\phi_n(\sigma_0)|^2  \sim 1/n^2 $.
Therefore, the behaviour of the series for $\tau \rightarrow 0{\scriptstyle
-} $ is

\begin{equation}
\begin{array}{l}
{\displaystyle \sum_{n \neq 0} | F_{\nu} \left( \frac{2\pi}{\sigma_0} |n|
\tau,
 \frac{2\pi}{\sigma_0} |n| \tau_0 \right) |^2 ~|\phi_n(\sigma_0)|^2
  \sim } \\ \\ {\displaystyle \sim
\frac{2^{2\nu}}{\Gamma^2(1-\nu)} \left( \frac{2\pi}{\sigma_0}
\right)^{2-2\nu} \! (-\tau_0)~ {-\tau}^{1-2\nu} \sum_{n \neq 0}^{}
\frac{ J_{\nu}^2 \left( -\frac{2\pi }{\sigma_0} |n|
\tau_0 \right) }{|n|^{2 \nu - 2}}~ |\phi_n(\sigma_0)|^2 }
\label{84}
\end{array}
\end{equation}

\noindent
Then, for $\nu \geq 1$ and $\tau_0 \rightarrow 0{\scriptstyle -}$ the
behaviour of $S_{II}$ is dominated by the upper limit of the integral. So,
inserting eq (\ref{84}) into eq. (\ref{73}) and taking into account eq.
(\ref{81}) we get

\begin{equation}
S_{II} \sim   \frac{\sigma_0}{\pi} \left( \frac{2\pi}{\sigma_0}
\right)^{2-2\nu} \! \frac{r^{2\nu - 2}}{J_{\nu}^2 (-r\tau_0)}~ \sum_{n \neq
0}^{} \frac{ J_{\nu}^2 \left( -\frac{2\pi }{\sigma_0} |n|
\tau_0 \right) }{|n|^{2 \nu - 2}} ~|\phi_n(\sigma_0)|^2 ~S_I
\label{85}
\end{equation}

\noindent
Now we take the long $\tau$ interval approximation by doing $\tau_0
\rightarrow -\infty$ in (\ref{85}). Thus

\begin{equation}
S_{II} \sim   \frac{\sigma_0}{\pi} \left( \frac{\sigma_0}{2\pi}
\right)^{2\nu -1} \sum_{n \neq 0}^{} \left( \frac{r}{|n|} \right)^{2\nu -1}
\frac{{\cos}^2 \left( - \frac{2\pi}{\sigma_0} |n| \tau_0 - \frac{\pi}{2}
\nu - \frac{\pi}{4} \right) }{ {\cos}^2 \left( - r\tau_0
- \frac{\pi}{2} \nu - \frac{\pi}{4} \right) }~ |\phi_n(\sigma_0)|^2 ~S_I
\label{86}
\end{equation}

\noindent
The factor

\noindent
\begin{equation}
\frac{1}{ {\cos}^2 \left( - r\tau_0
- \frac{\pi}{2} \nu - \frac{\pi}{4} \right) }
\label{87}
\end{equation}

\noindent
comes from the reciprocal of the Bessel functions $ J_{\nu}^2 (-r\tau_0)$
for $\tau_0 \rightarrow -\infty$, entering in (\ref{85}), and which was
assumed not to vanish. In particular we can choose $\tau_0$ in such a way
that the factor (\ref{87}) is $1$. This yields an upper bound estimate of
the action $S_{II}$ of the form

\begin{equation}
S_{II} \leq \left( \frac{\sigma_0}{2\pi} \right)^{2\nu -1}
\frac{\sigma_0}{\pi}~  \sum_{n \neq 0}^{} \left( \frac{r}{|n|}
\right)^{2\nu -1}| \! \phi_n(\sigma_0)|^2 ~S_I
\label{88}
\end{equation}

\noindent
and for string $III$ we have

\begin{equation}
S_{III} \leq \left( \frac{2\pi - \sigma_0}{2\pi} \right)^{2\nu -1}
\frac{2\pi - \sigma_0}{\pi}~  \sum_{n \neq 0}^{} \left( \frac{r}{|n|}
\right)^{2\nu -1} \! |\phi_n(2\pi -\sigma_0)|^2 ~S_I
\label{89}
\end{equation}

\noindent
Notice that this upper bound becomes exact for $\sigma_0 = 0$ and
$\sigma_0 = 2\pi$ (the non-splitting solution).

In order to get a better insight on the behaviour of
$\Delta S$, we choose $\alpha =2$ ($\nu = 3/2$), in which case, the series
(\ref{88}) and (\ref{89}) can be sumed in closed form. For $\alpha = 2$
we have

\begin{equation}
\begin{array}{lcl}
S_{II}^{(\alpha =2)} & \leq & {\displaystyle \frac{\sigma_0}{2\pi} \left( 1+
\frac{\sin 2r\sigma_0}{2r\sigma_0} - \frac{ 2\, {\sin}^2 r\sigma_0 }{ r^2
\sigma_0^2} + \right.Ê}
\\    & & \\   & & {\displaystyle \left. + \frac{{\sin}^2 r\sigma_0}{6} +
\frac{2\, \sin r\sigma_0}{r\sigma_0}  - \frac{ 8\, {\sin}^2\,
\frac{1}{2}r\sigma_0 }{ r^2 \sigma_0^2}\right) S_I  }
\end{array}
\label{90}
\end{equation}

\noindent
Thus

\begin{equation}
\begin{array}{l}
\Delta S^{(\alpha =2)}  = ( S_{II}^{(\alpha =2)} +
S_{III}^{(\alpha =2)}) - S_{I}^{(\alpha =2)} \leq \\
 \\
{\displaystyle \leq
\left\{ - \frac{1}{\pi r^2} \left( \frac{1}{\sigma_0} +
\frac{1}{2\pi - \sigma_0} \right) {\sin}^2 r\sigma_0 + \frac{{\sin}^2
r\sigma_0}{6}
  - \frac{4}{\pi r^2} \left( \frac{1}{\sigma_0} +
\frac{1}{2\pi - \sigma_0} \right) {\sin}^2\, \frac{r\sigma_0}{2}
 \right\} S_I }
\end{array}
\label{91}
\end{equation}

For the fundamental case $r=1$, the action difference takes the form

\begin{equation}
\Delta S_1^{(\alpha =2)} = \left\{
\left[ \frac{1}{6} - \frac{1}{\pi} \left( \frac{1}{\sigma_0} +
\frac{1}{2\pi - \sigma_0} \right) \right] {\sin}^2 \sigma_0
  - \frac{4}{\pi } \left( \frac{1}{\sigma_0} +
\frac{1}{2\pi - \sigma_0} \right) {\sin}^2\, \frac{\sigma_0}{2}
\right\} S_{1,I}
\label{92}
\end{equation}

\noindent
which is a negative quantity for all values of $\sigma_0$ in the interval
$[0,2\pi]$. Thus, in the background of the singular gravitational plane
wave ($\alpha =2$), the action of the splitting solution is smaller than the
action of the non-splitting one. Moreover, this effect is magnified by an
overall divergent $\tau_f$ dependent factor when we approach the singularity
at $\tau =0$, because the action $S_I^{(\alpha =2)}$ is multiplied by such
a factor in this limit (eq.(\ref{81}). In addition, the effect of smaller
action for the splitting solution is also increased in relative terms, as a
consequence of the new terms appearing in (\ref{92}). This is easily seen in
terms of the lower bound that we have for the relative decrease in area.
According to (\ref{92})

\begin{equation}
\eta_1^{(\alpha =2)}(\sigma_0) =
\frac{|\Delta S_1^{(\alpha =2)}|}{S_{1,I}^{(\alpha =2)}} \geq h(\sigma_0)
\label{93}
\end{equation}

\noindent
where

\begin{equation}
h(\sigma_0) = \left[   \frac{1}{\pi} \left( \frac{1}{\sigma_0} +
\frac{1}{2\pi - \sigma_0} \right) - \frac{1}{6} \right] {\sin}^2 \sigma_0
  + \frac{4}{\pi } \left( \frac{1}{\sigma_0} +
\frac{1}{2\pi - \sigma_0} \right) {\sin}^2\, \frac{\sigma_0}{2}
\label{94}
\end{equation}

\noindent
The lower bound $h(\sigma_0)$ has its maximum at $\sigma_0 = \pi$ with a
heigth $h_{\rm max} = 0.811$. Therefore

\begin{equation}
\eta_{1,{\rm max}}^{(\alpha =2)} > 0.811
\label{95}
\end{equation}

\noindent
which is much larger than the relative decrease in area for the same
string configuration in flat space-time, given in (\ref{eta2max}).

\section{Concluding remarks}

In this paper we have explored a new type of solutions of the string
equations of motion and constraints. These solutions describe the splitting
of strings as a natural decay process that takes place in real
\mbox{(lorentzian signature)} space-time. This process occurs at the
classical level and this is natural because the string is an extended
object. The splitting solutions are already present in flat space-time, and
they correspond to stationary points of the action (area) with lower value
than the non splitting strings. In order to explore the effect of a
gravitational field
on the splitting solutions, we have considered a gravitational singular plane
wave background. In the case that we analyze, the gravitational field
produces an enhacement of the effect of smaller area for the splitting
solution. It would be interesting to settle to what extent these results are
universal, and work in this direction is now in progress by the present
authors.

   On the other hand, string splitting is usually considered and discussed
within a quantum formulation, namely the euclidean path integral functional
approach to the quantum string scattering amplitudes. In this context,
the stationary points  of the euclidean action correspond to solutions of
Dirichlet or Neumann boundary value problems for elliptic operators
(laplacians)  on bordered Riemann surfaces \cite{accion}, i.e. the
classical string equations of motion and constraints lead to solve boundary
value problems for elliptic operators with Dirichlet or Neumann boundary
conditions. Of course, these are different from the solutions considered in
this paper, in which we solve the hyperbolic (lorentzian) evolution
equations for the Cauchy data $X^{A}(\sigma,\tau_0)$ and
${\dot{X}}^A(\sigma,\tau_0)$ with some fixed topology. This topology should
be viewed as enforced by the world-sheet metric used to construct the
D'Alambertian operator. Notice that quantum mechanically the initial data
$X^{A}(\sigma,\tau_0)$ and ${\dot{X}}^A (\sigma,\tau_0)$ can not be given
simultaneously. Instead, one gives the initial and final string shapes to
compute a transition amplitude between them.

Although our splitting solutions are purely classical, string splitting for
 massive strings is also present at the quantum level. The relevant
magnitude to be computed in that case is the probability amplitude for such
a process. In fact, such probability has been computed in \cite{Polch} for
flat space-time. It would be very interesting to explore the relationships
between the classical splitting solutions and the quantum probability for
string desintegration, and also the effect of a gravitational field on such
probability. The quantum probability amplitude for string splitting in a
singular plane wave will be discussed by the present authors in a
forthcoming paper.

We conclude with a  final remark concerning the classical solutions of the
string equations of motion in curved spaces-times. It has been stablished
(see for instance \cite{Erice} and references therein) that  these solutions
present a phenomenon of indefinitely string stretching near space-time
singularities, due
to the absorption by the string of energy from the background gravitational
field. Of course, there must be a mechanism that avoids this indefinite string
growing, and indeed  the strings can radiate away energy by
emiting gravitons or other particle like excitations. However,  another
natural mechanism to avoid string growing is string splitting, and it
would be interesting to elucidate its quantitative relevance to avoid
the  indefinite stretching of strings in strong gravitational fields.


{\bf Acknowledgements}

One the authors (J.R.M.) acknowledges the hospitality at LPTHE (Universit\'e
de Paris VI-VII) and at DEMIRM (Observatoire de Paris)
where part of this work was carried out. Also J.R M.
acknowledges the Direcci\'on General de Investigaci\'on Cient\'{\i}fica y
T\'ecnica (DGICYT) for financial support.


\newpage

\begin{thebibliography}{99}
\bibitem{Witten} See for example: \\
M. Green, J. Schwarz, E. Witten, {\em Superstring Theory
\/}. Cambridge University Press. 1987.
\bibitem{Poly} A. M. Polyakov, {\em Phys. Lett. \/} {\bf 103B}, 207
(1981).
\bibitem{onda} H.J. de Vega and N. S\'anchez, {\em Phys. Rev. \/}
{\bf D45}, 2783 (1992). \\
 H.J. de Vega,  M. Ram\'on Medrano and N. S\'anchez, \\
{\em Class. and Quantum Gravity\/} {\bf  10}, 2007 (1993).

\bibitem{Polch} J. Dai and J. Polchinski, {\em Phys. Lett. \/}
{\bf 220B}, 387 (1989). \\ R.B. Wilkinson, N. Turok and D. Mitchell,
{\em Nuclear Physics \/}  {\bf B332}, 131 (1990).

\bibitem{accion} J. Ram\'{\i}rez Mittelbrunn and M.A. Mart\'{\i}n Delgado,\\
{\em Int. Journal of Mod. Phys.}, {\bf  A 6}, 1719 (1991).

\bibitem{Erice} H.J. de Vega and N. S\'anchez in {\em String Quantum
Gravity and Physics at the Planck Energy Scale} Erice, 1992. World
Scientific.

\end{thebibliography}
\end{document}